\PassOptionsToPackage{pdfpagelabels=false}{hyperref}

\documentclass[fleqn,usenatbib]{mnras}

\usepackage{newtxtext,newtxmath}

\usepackage[T1]{fontenc}
\usepackage{ae,aecompl}
\usepackage{soul}

\usepackage{graphicx}   
\usepackage{amsmath}    
\usepackage{amssymb}    

\usepackage{multirow}
\usepackage{deluxetable}

\title[A Study of Single Pulses in the Parkes Multibeam Pulsar Survey]{A Study of Single Pulses in the Parkes Multibeam Pulsar Survey}

\author[M. B. Mickaliger et al.]{
Mitchell B. Mickaliger,$^{1,2}$\thanks{E-mail: mitchell.mickaliger@manchester.ac.uk (MBM)}
A. E. McEwen,$^{2,3}$
M. A. McLaughlin$^{2,4}$ and
D. R. Lorimer$^{2,4}$
\\
$^{1}$Jodrell Bank Centre for Astrophysics, School of Physics and Astronomy, The University of Manchester, Manchester M13 9PL, UK\\
$^{2}$Department of Physics and Astronomy, West Virginia University, Morgantown, WV 26506, USA\\
$^{3}$Department of Physics and Astronomy, University of Wisconsin-Milwaukee, Milwaukee, WI 53211, USA\\
$^{4}$Center for Gravitational Waves and Cosmology, Chestnut Ridge Research Building, Morgantown, WV 26505, USA\\
}

\date{Accepted XXX. Received YYY; in original form ZZZ}

\pubyear{2017}

\begin{document}
\label{firstpage}
\pagerange{\pageref{firstpage}--\pageref{lastpage}}
\maketitle

\begin{abstract}
We reprocessed the Parkes Multibeam Pulsar Survey, searching for single pulses out to a DM of 5000~pc cm$^{-3}$ with widths of up to one second. We recorded single pulses from 264 known pulsars and
14 Rotating Radio Transients. We produced amplitude distributions for each pulsar which we fit with log-normal distributions, power-law tails, and a power-law function divided by an exponential function, finding that some pulsars show a deviation from a log-normal distribution in the form of an excess of high-energy pulses. We found that a function consisting of a power-law divided by an exponential fit the distributions of most pulsars better than either log-normal or power-law functions. For pulsars that were detected in a periodicity search, we computed the ratio of their single-pulse signal-to-noise ratios to their signal-to-noise ratios from a Fourier transform and looked for correlations between this ratio and physical parameters of the pulsars. The only correlation found is the expected relationship between this ratio and the spin period. Fitting log-normal distributions to the amplitudes of pulses from RRATs showed similar behaviour for most RRATs. Here, however, there seem to be two distinct distributions of pulses, with the lower-energy distribution being consistent with noise. Pulse-energy distributions for two of the RRATS processed were consistent with those found for normal pulsars, suggesting that pulsars and RRATs have a common emission mechanism, but other factors influence the specific emission properties of each source class.
\end{abstract}




\section{Introduction}
\label{intro}

While pulsars are widely used for testing relativity \citep[e.g.,][]{Taylor:1982, Kramer:2006, Antoniadis:2013} and frequently timed for use in pulsar timing arrays \citep[e.g.,][]{Jenet:2009, Ferdman:2010, Hobbs:2010, Manchester:2013}, their emission processes \citep[see, e.g.,][]{Weatherall:1998, Istomin:2004, Petrova:2006} and equation of state \citep[e.g.,][]{Demorest:2010, Katayama:2012} are not fully understood. Further insights from phenomenological studies of single pulses are particularly useful to inform theoretical models.

Pulsars are most commonly detected through periodicity searches, though some of them are also detectable through their single pulses. One class of pulsars, the Rotating Radio Transients (RRATs), are detected more strongly in single pulses than periodicity searches \citep{Keane:2011}. By studying these single pulses, we can constrain the pulsar emission mechanism. Previous studies have shown that there seem to be two distinct categories of single pulses: normal pulses and `giant'-pulses \citep[see, e.g.,][]{Lundgren:1995, Johnston:2002, Kramer:2002}. Although giant pulses are sometimes arbitrarily defined as pulses of more than 10 times the average pulse energy \citep[e.g.,][]{Karuppusamy:2010, Karuppusamy:2012}, a comparison of the energies of normal pulses and giant pulses none the less show that normal pulses follow a log-normal distribution \citep{Cognard:1996, Cairns:2001, Cairns:2004}, whereas giant pulses have energies that fit a power-law distribution \citep{Lundgren:1995, Johnston:2002, Kramer:2002}. This suggests that the emission mechanisms for these two types of pulses are different. \citet{Knight:2006} argue that the definition of a giant pulse should therefore be based on short time-scale, narrow-phase emission that has power-law energy statistics. Given either definition, however, pulses with amplitudes between those of normal pulses and giant pulses have yet to be fully studied. \citet{Lundgren:1995} have seen an intrinsic rollover in the energy distribution of giant pulses at low energies, suggesting that there is a gap between normal pulses and giant pulses. However, observations have not yet ruled out the possibility that the power-law distribution of giant pulses is merely a high-energy tail on a log-normal distribution of normal pulses \citep{Karuppusamy:2011}. These energy distributions can provide insight into the pulsar emission mechanism, so determining their true nature is important. In order to successfully do this, we need a representative sample of single pulses from a large number of objects. The number and variety of pulsars present in pulsar surveys, like the Parkes Multibeam Pulsar Survey \citep[hereafter PMPS;][]{Manchester:2001}, provides us with the opportunity to carry out such a study.

One of many ways with which we can constrain single-pulse emission is by searching for
correlations between pulse strength and the physical properties of these pulsars. In our analysis, 
we quantify pulse intensity and compare it to the pulsar's spin period ($P$), period derivative 
($\dot{P}$), dispersion measure (DM), characteristic age ($\tau_{\rm c}$), spin-down energy loss 
rate ($\dot{E}$), surface magnetic field ($B_{\rm surf}$), and magnetic field strength at the 
light cylinder ($B_{\rm LC}$). A similar analysis by \citet{Cui:2017} found that the period 
derivatives of RRATs were slightly larger than those of other pulsars, hinting at a possible 
correlation. \citet{Cui:2017} also found that the amplitude distributions of RRATs are log-normal, 
with some showing a power-law tail.

A larger study of this kind has been performed by \citet{Burke-Spolaor:2012}, who investigated single pulses detected in the High Time Resolution Universe (HTRU) survey \citep{Keith:2010}. They detected
single pulses from 315 known pulsars and performed energy distribution fits using both log-normal
and Gaussian distributions, finding that most of the pulsars fit log-normal energy distributions,
with only a few favouring a Gaussian distribution. They found that some of the energy distributions
had multiple peaks, which they showed to be caused by mode changes in these pulsars. These mode
changes, which may be due to magnetospheric reconfigurations leading to weaker or redirected
emission \citep{Timokhin:2010}, can account for the nulls seen in some pulsars. They also found no
correlation between modulation parameters and physical properties.

While their study covers a similar region of sky to ours, and thus analyses many of the same sources, \citet{Burke-Spolaor:2012} did not compare detectability in 
single-pulse and periodicity searches as we do in this paper. In addition, the PMPS integration 
time is longer, allowing us to record more single pulses, putting better constraints on our amplitude distributions. Also, we searched for single pulses with widths of up to one second, while \citet{Burke-Spolaor:2012} were only sensitive to pulses with widths less than 32~ms.

Here we present an analysis of single pulses from all known pulsars and RRATs detected in the
PMPS. In \S \ref{data} we describe the data reduction and analysis. \S \ref{known}
discusses all of the known pulsars from which we detected single pulses, while \S \ref{rrats}
presents the single-pulse results from the RRATs. Finally, conclusions are given in \S \ref{conc}.

\section{Data Reduction}
\label{data}

We used freely-available software tools implemented in the {\tt sigproc} software 
package\footnote{\url{http://sigproc.sourceforge.net}} to search the PMPS data for periodic and single-pulse sources. The results of our periodicity search are presented in \cite{Mickaliger:2012}. Here we focus on the search for individual pulses. We searched each beam in the survey for both periodicity candidates and single pulses out to a DM of 5000~pc cm$^{-3}$, selecting events with signal-to-noise ratio (S/N) above 5 and with 
sensitivity to events with widths ranging from the sampling time of 250 $\mu$s to 1~s to search for fast radio bursts (FRBs). No new FRBs were found, but in the process we recorded single pulses from known pulsars in order to exclude them from our FRB candidates.

\begin{figure*}
\centering
\includegraphics[height=\textwidth, angle=-90]{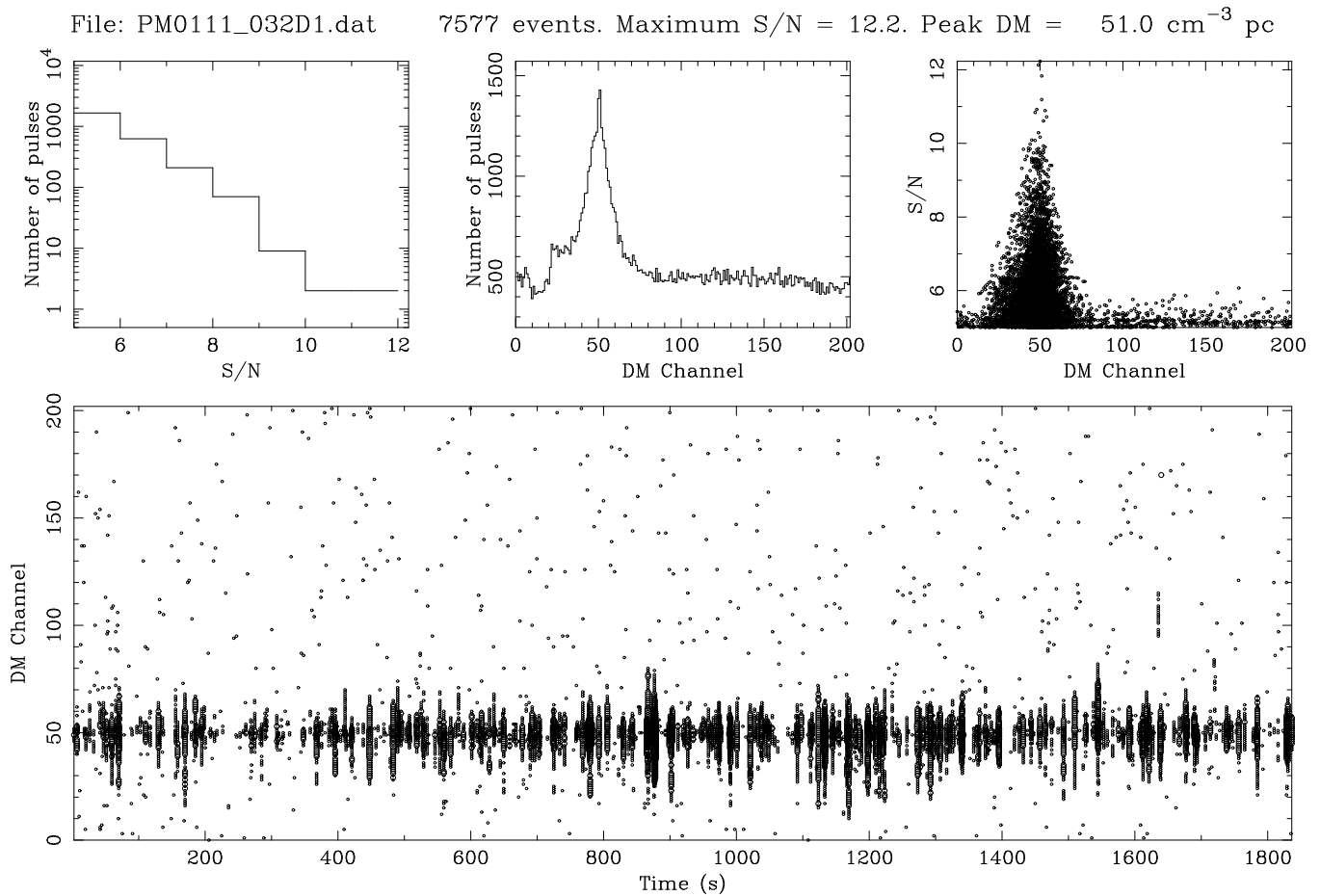}
\caption[Example of Single-Pulse Search Diagnostic Plot]{Example of our single-pulse analysis for 
PSR J0924$-$5814. The top plots, from left to right, show the number of events versus S/N, number 
of events versus DM channel, and S/N versus DM channel. The middle and right plots show an excess 
of events centred at a DM channel of $\sim$50 (corresponding to a DM of $\sim$60~pc 
cm$^{-3}$). The arrival time (relative to the start of the observation) and DM of each detected 
event can be seen in the bottom plot. The size of each point on the plot corresponds to the S/N of 
the respective event.}
\label{fig:single}
\end{figure*}

Initially, frequency channels in the data were shifted to correct for dispersion due to free
electrons in the interstellar medium, a process called dedispersion. The amount the channels
were shifted was based on the DM searched. To search for pulsars and single pulses, we dedispersed
the data using many different DM values. The total number of DMs searched was 203 and was
optimally chosen by {\tt dedisperse\_all}\footnote{\url{http://www.github.com/swinlegion}}, the
program we used to dedisperse our data due to its speed and efficiency. Dedispersion of the data
led to a time series for each DM, and these time series were then processed by {\tt seek}, a
program which searches for periodic and single-pulse signals from a source. The periodicity-search
analysis implemented in {\tt seek} is the standard Fourier-based approach where the amplitude
spectrum is subject to multiple harmonic folds, summing 2, 4, 8, and 16 harmonics. This process increases sensitivity to narrow pulses in a close-to-optimal fashion \citep{Ransom:2002}. The single-pulse analysis performed by {\tt seek} uses a top-hat function to match-filter events of different widths. In order to be sensitive to wider pulses, the time samples in the data are smoothed by factors of two, i.e., the data are added in pairs at each smoothing step. In our analysis, we used a total of 12 smoothings, resulting in a search for pulses with widths of up to $2^{12}$ time samples (1.024~s). There is no sensitivity to wider pulses due to the 0.9~s high-pass filter time constant used 
in PMPS data taking \citep{Manchester:2001}. Once all of the DMs were searched, all events with S/N $>$ 5 were saved. These events are used to create a single-pulse plot (see Fig. \ref{fig:single} for an example), which shows the arrival time of each event versus its DM, as well as the S/N of each event.

For each pulsar in the ATNF pulsar
catalogue\footnote{\url{http://www.atnf.csiro.au/people/pulsar/psrcat}}, we examined candidates from a periodicity search of the closest beam(s). We
kept all periodicity candidates with an FFT S/N $>$ 9. If the spin period and DM of the pulsar
were within 10 per cent of the spin period (or its harmonics) and DM of any of the periodicity candidates, we noted 
the pulsar as being detected in a periodicity search of that beam, and also recorded its FFT S/N. 
For pulsars which were detected in a periodicity search of equidistant beams, we used the beam in 
which the FFT S/N was highest for our analysis. We inspected the single-pulse plot from the beam 
by eye to determine if single pulses were visible, in which case we selected all events with S/N 
$>$ 5 and within 10 per cent of the DM of the pulsar. We then folded these candidate pulses using parameters obtained from the ATNF pulsar catalogue. Candidates that were not in phase with the majority of the candidates were deemed spurious and removed.

For each pulsar, we then identified all beams
for which the beam centre was within 30 arcmin of the pulsar. We searched the events recorded from these beams for candidates within 10 per cent of the DM of the pulsar. We assumed that
events with a DM $<$ 1~pc cm$^{-3}$ were radio frequency interference (RFI). To determine
if events at other DMs were associated with RFI, we measured the width and the arrival time
of each remaining event, that is, any event not already classified as a candidate or zero-DM RFI. If the arrival time of a candidate was within the pulse width of the arrival time of an event, we compared their S/Ns. If the S/N was higher in the event, we considered the candidate and the event to be RFI. Otherwise, we considered the event to be a candidate single pulse related to the pulsar. We performed the same reduction for all known RRATs\footnote{\url{http://astro.phys.wvu.edu/rratalog}}. Events that were not attributed to a known pulsar, RRAT, or RFI were recorded as burst candidates. We then closely inspected all burst candidates and determined them to be either RFI or recently discovered astrophysical sources.

Some single pulses are wide enough that they are detected in consecutive time samples, in which case we would record each event in these consecutive time samples as a candidate. Since these multiple candidates are from one single astrophysical pulse, we need to account for this when determining the number of single pulses recorded from a source. For every candidate recorded
within one spin period, we compared it with the preceding and subsequent candidate within that spin period. If it was not the strongest of the three, the next candidate was tested. After recording the strongest candidate as the peak a single astrophysical pulse (and attributing the other candidates to this pulse as well), we determined the number of single pulses seen in each observation and the maximum single-pulse S/N.

We also calculated the energy of every pulse in order to create pulse-energy distributions. To do this, we dedispersed the data from a beam at the DM of the pulsar in that beam, using the DM value published in the ATNF pulsar catalogue. We then folded the time series at the pulse period and computed the on-pulse energy in every rotation by integrating over a number of phase bins in an on-pulse window and subtracting the off-pulse mean. Finally, we normalised the on-pulse energies by dividing them by the mean on-pulse energy over all integrations. The on-pulse window differed for each pulsar, and was chosen to encompass the total folded profile, based on visual inspection. Since there was significant deviation in the number of noise bins selected when determining the on-pulse window by hand, we tested the sensitivity of the resulting amplitude distribution to the inclusion of many noise bins in the on-pulse window. We found that there was no significant change to the amplitude distributions for a wide range of noise bins included in on-pulse windows. To calculate the off-pulse mean, we first took the average of each off-pulse window on either side on our on-pulse window. We took the difference of these averages and multiplied them by the pulse width. We then subtracted this area from our total folded profile.

\section{Known Pulsars}
\label{known}

In total, we recorded single pulses from 264 known pulsars out of the 1049 present in the survey.
We based our detections on pulses from the beam closest to each of the pulsars since we could
not reliably scale the energies of pulses that were substantially offset from the centre of the
beam. For each detected pulsar, we constructed a normalised energy distribution, which we then fit
with log-normal and power-law distributions, as well as a power-law divided by an exponential \footnote{Plots of each fit (log-normal, power-law, power-law divided by exponential) to the energy distributions of every pulsar can be found at \url{http://astro.phys.wvu.edu/pmps/single.html}}. We calculated the $(S/N_{\rm SP})/(S/N_{\rm FFT})$ ratio and compared this to the physical properties of the pulsar (spin period, period derivative, DM, characteristic age, spin-down energy loss rate, surface magnetic field strength, and magnetic field strength at the light cylinder) to determine if there is a correlation between them.

\subsection{Energy Distributions}
\label{dists}

\begin{figure*}
\centering
\includegraphics[height=\textwidth, angle=-90]{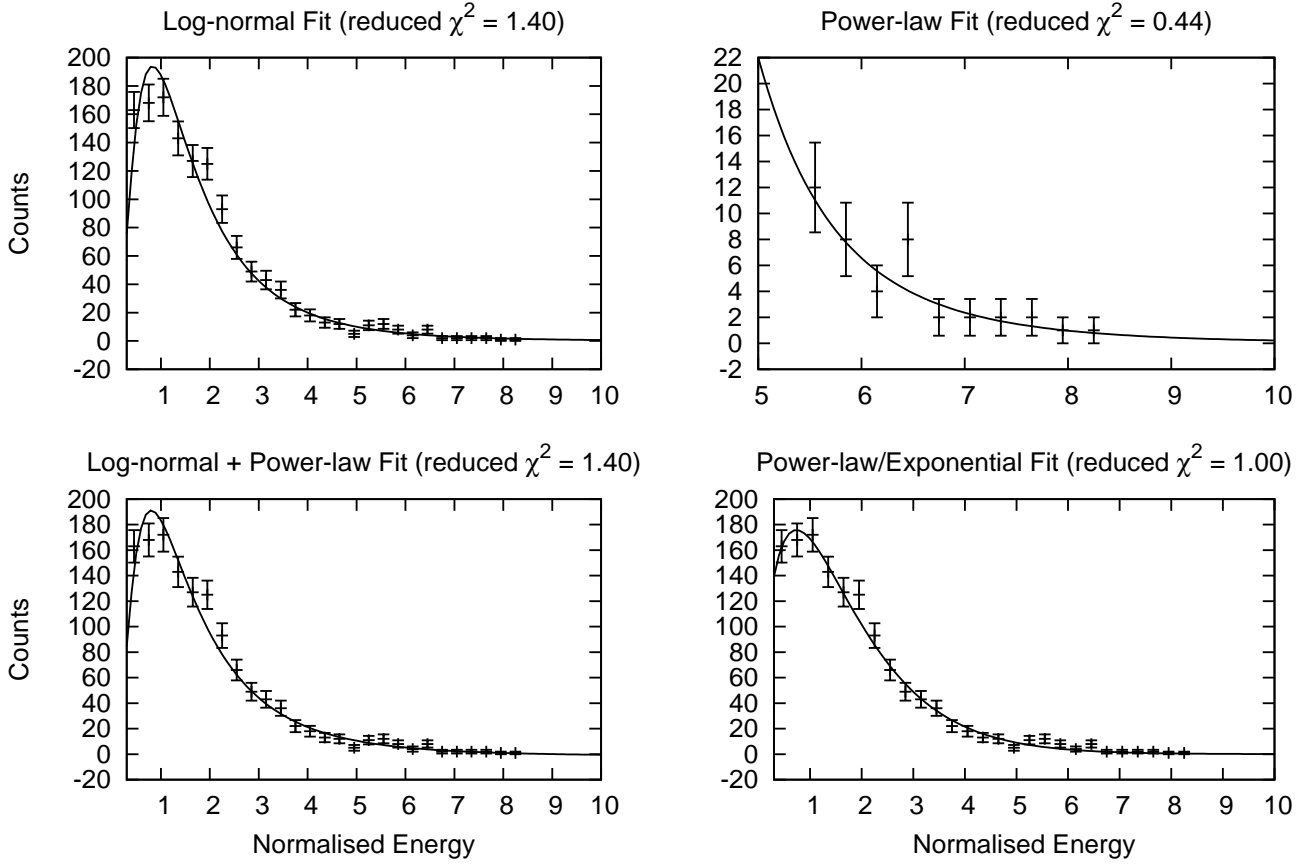}
\caption[Comparison of Fitting Functions]{Amplitude distributions with the four fitting functions
compared in this paper. The pulsar used for this comparison is PSR J0955$-$5304. Clockwise from the
top left, the fitting functions are log-normal ($\chi_{\rm red}^2$ = 1.40), power-law tail ($\chi_{\rm red}^2$ = 0.44), power-law$/$exponential ($\chi_{\rm red}^2$ = 1.00), and log-normal plus power-law ($\chi_{\rm red}^2$ = 1.40). Note that the power-law tail fit is displayed over the portion of the amplitude distribution to which it is fit, i.e. the ten highest energy bins.}
\label{fig:fit_comparison}
\end{figure*}

\begin{figure*}
\centering
\includegraphics[height=\textwidth, angle=-90]{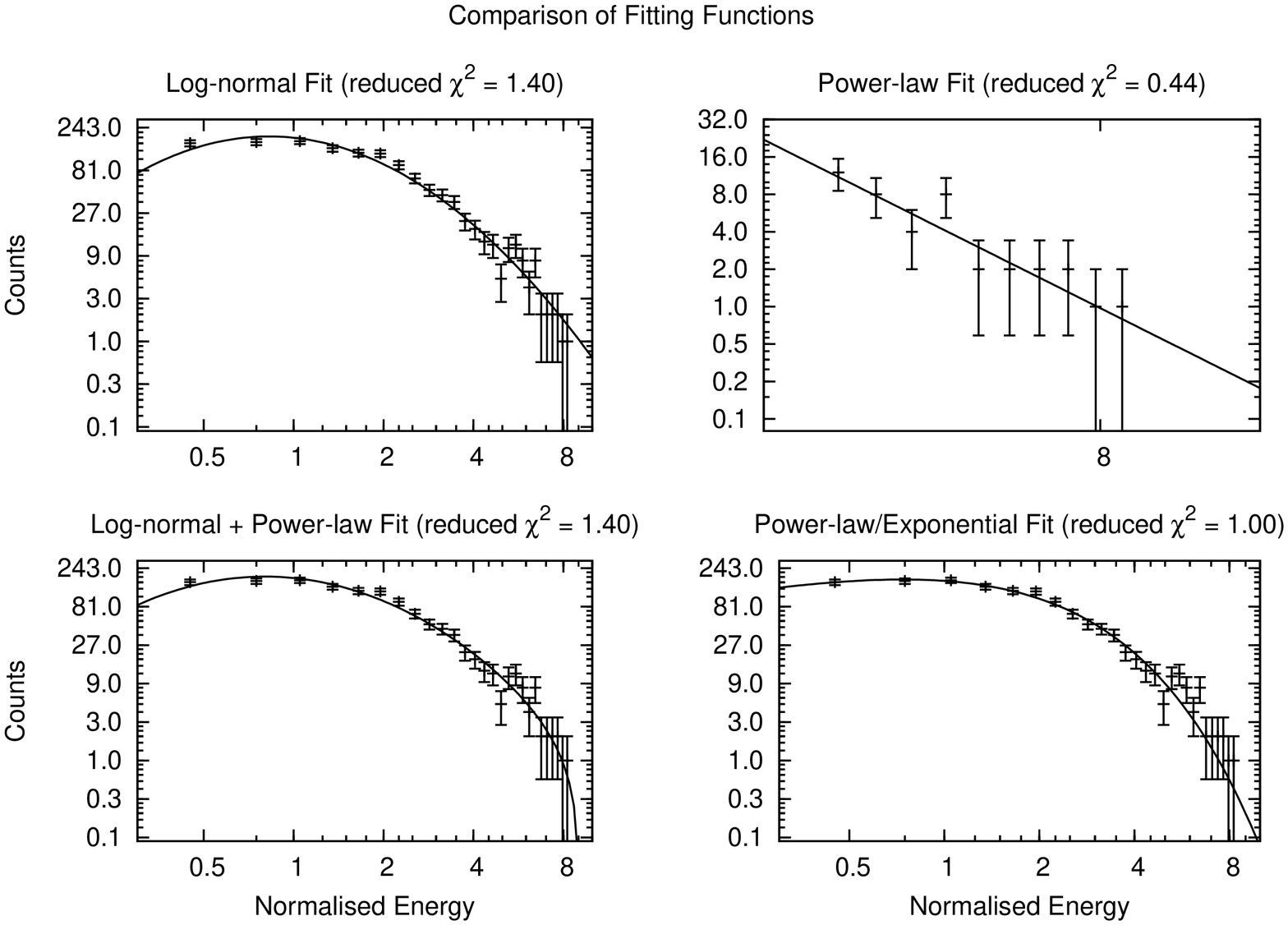}
\caption[Comparison of Fitting Functions (log-normal)]{Log-normal axes representation of the amplitude distributions with the four fitting functions compared in this paper. The pulsar and functions are the same as those in Fig. \ref{fig:fit_comparison}.}
\label{fig:fit_comparison_log}
\end{figure*}

In order to create energy distributions for each pulsar, we used the standard method of
normalising the energy of each pulse by dividing it by the average energy over the entire
observation \citep{Ritchings:1976, Biggs:1992, Burke-Spolaor:2012}. We binned these normalised
energies into differential distributions which we then fit, through least-squares fitting, with
log-normal distributions and power-law tails. Our log-normal fit has the form of a scaled
probability density function (PDF), given by
\begin{equation}
{\rm PDF}_{\rm log}(E)=\frac{C}{E\sigma\sqrt{2\pi}} \exp 
\left[ -\frac{(\ln{E}-\mu)^2}{2\sigma^2}\right],
\label{eq:lognormal}
\end{equation}
where $E$ is the normalised energy, $C$ is a scaling factor, $\mu$ is the mean normalised energy,
and $\sigma$ is the standard deviation. Our power-law fit has the form
\begin{equation}
{\rm PDF}_{\rm power}(E)=CE^{\alpha},
\label{eq:power}
\end{equation}
where $E$ is the normalised energy, $C$ is a scaling factor, and $\alpha$ is the power-law index. 
The sum of these two PDFs was also fit against the energy distributions, but did not fit as well 
as the other functions mentioned. For this reason, the fit parameters for this summed function are not included in this paper. Plots of the power-law plus log-normal functions are included in Figs. \ref{fig:fit_comparison} and \ref{fig:fit_comparison_log}.

Following these fits, it became clear that in many cases, the log-normal fits did not fit the energy
distributions as closely as we expected (i.e. the $\chi_{\rm red}^2$ was greater than 1, often
significantly). In most cases, the number of higher energy pulses was lower than predicted 
from the best-fitting log-normal distribution. To address this, a new PDF with a steeper decay at 
higher energies was fit to the data. This was defined as 
\begin{equation}
{\rm PDF}_{\rm pow/exp}(E)=\frac{aE^{b}}{e^{cE}},
\label{eq:powexp}
\end{equation}
where $a$, $b$, and $c$ are fitting parameters.

The best-fitting parameters for each pulsar, as well as the $\chi_{\rm red}^2$ of the fit, the spin
period, DM, number of single pulses recorded, maximum single-pulse S/N, FFT S/N (if the pulsar was
detected in a periodicity search in the closest beam), and the $(S/N_{\rm SP})/(S/N_{\rm FFT})$
ratio are listed in Table \ref{table:energy}. The errors reported for the fit parameters represent
one sigma errors reported by the least-squares fitting.

In pulse periods where no pulse is detected, the values of the pulse energy can be zero, positive,
or negative. Since there is no physical meaning to a negative pulse energy (and these negative
values are likely due to random noise fluctuations), we set these energies to zero. We can then
identify nulling pulsars by looking for a large excess of zero-energy pulses in the pulse-energy 
distributions.

\begin{figure*}
\centering
\includegraphics[height=\textwidth, angle=-90]{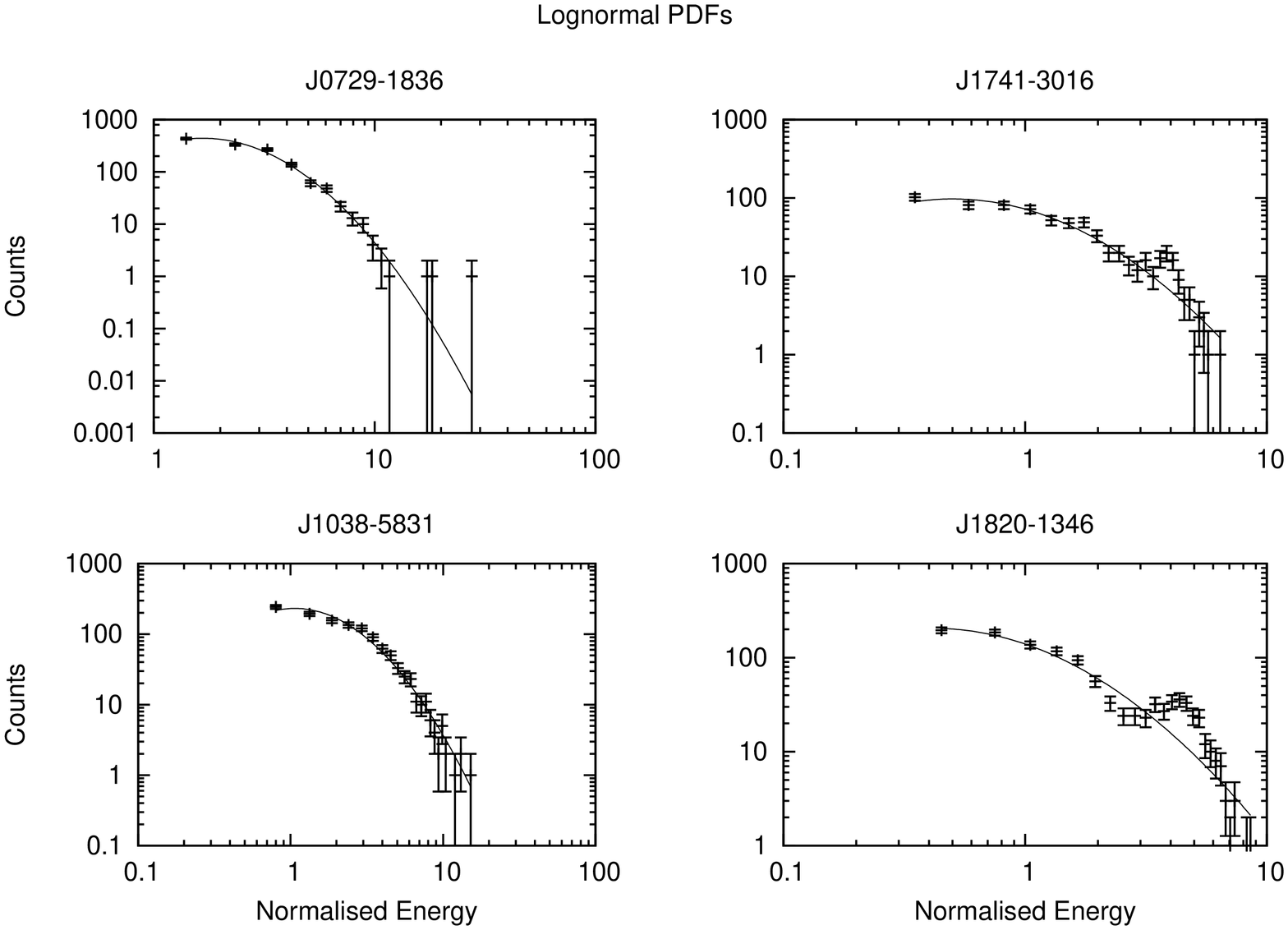}
\caption[Examples of Log-normal Fits to Amplitude Distributions]{Amplitude distributions with 
log-normal fits. The two plots on the left are examples of normal pulsars, while the two plots on 
the right show `bumps' in their distributions that indicate bursting pulsars. Here, normalised
energy is defined as the individual pulse energy divided by the average energy of all pulses. The 
pulsars are, clockwise from top left, PSRs J0729$-$1836, J1741$-$3016, J1820$-$1346, and 
J1038$-$5831.}
\label{fig:burst}
\end{figure*}

Some pulsars display `bumps' on their energy distributions, and we can use these excess pulses to 
determine the average S/N while the pulsar is `bursting,' and also to estimate how many pulses the 
pulsar emits in this bursting state. A comparison of the amplitude distributions of these bursting 
pulsars to normal pulsars can be seen in Fig. \ref{fig:burst}. An inspection of the amplitude 
distributions show that roughly 17 pulsars display this bursting behaviour.

As can be seen in Table \ref{table:energy}, there is a wide range for estimates of $\mu$ in our log-normal fits, ranging from $\sim-$11 $E/\langle{E}\rangle$ to $\sim$3.1 $E/\langle{E}\rangle$. A negative value of $\mu$ is not physically possible, since no pulses have negative energy, but the nine measurements with a negative value of $\mu$ have estimated standard deviations ($\sigma$) that put them within 1$\sigma$ of zero. The range of $\sigma$ values is $-$0.85 $E/\langle{E}\rangle$ to         
$\sim$3.2 $E/\langle{E}\rangle$, but the large estimated standard deviation for the upper 
measurement make $\sim$1.5 $E/\langle{E}\rangle$ a more likely upper limit. The ranges for both of 
these parameters are larger than those seen in previous studies (i.e. \citet{Burke-Spolaor:2012}), 
but are still consistent with those results. The average $\chi_{\rm red}^2$ for these fits is 3.18.

For some pulsars, the tails of their distributions are well fit by a power-law. As previously 
stated, energy distributions of giant pulses are well represented by a power-law, whereas normal 
pulses are better fit by a log-normal. The power-law indices from these fits range from $-$15 to 
2. A visual inspection of the amplitude distributions shows that $\sim$30 pulsars have power-law tails, as defined by $0.90 < \chi_{\rm red}^2 < 1.10$.                              

The fit parameters for the power-law/exponential generally fit the energy distributions best,
and the ranges of the parameters are narrower. The values of $b$ range from $-$1.4 to 2.83,
with most values ranging between $\sim$0.3 and $\sim$1.2. The $c$ values range from $-$0.06 to 
2.25, with the majority of values between $\sim$0.5 and $\sim$1.1. The $\chi_{\rm red}^2$ are also
in general much closer to 1, with an average value of 1.61. This function also fit more of the 
distributions, with only 21 distributions that were unable to be fit as compared to the 33 for 
which a log-normal fit could not be found.

\subsection{Comparison of \texorpdfstring{$(S/N_{\rm SP})/(S/N_{\rm FFT})$}{(SP S/N)/(FFT S/N)} Ratio to Physical Properties}
\label{sp_comp}

The $(S/N_{\rm SP})/(S/N_{\rm FFT})$ ratios we calculated range from 0.01 to 1.51. This ratio has previously been used by \citet{McLaughlin:2003}, who also defined it as
\begin{equation}
r=\frac{2\eta}{\zeta{N_{\rm p}^{1/2}}}\frac{S_{\rm max}}{S_{\rm av}^{\prime}},
\label{eq:intermitt}
\end{equation}
where $\eta$ and $\zeta$ are pulse-shape dependent factors ($\eta\sim$1 and $\zeta\approx$1.06 for a Gaussian pulse), the number of pulses $N_{\rm p}=T_{\rm obs}/P$, $S_{\rm max}$ is the S/N of the pulse peak, and $S_{\rm av}^{\prime}$ is the modified mean intensity. This `modified' mean intensity refers to the average pulse intensity after renormalisation, defined by \citet{McLaughlin:2003} as 
\begin{equation}
S_{\rm av}^{\prime}=S_{\rm av}\left[{1-\frac{\log{N_{\rm p}}}{(N_{\rm p}-1)}}\right].
\label{eq:s_prime}
\end{equation}
\citet{Deneva:2009} used this definition to assign $r$ values to pulsars detected in both the PMPS and PALFA surveys, finding $0.005 \lesssim r \lesssim 1$ for PMPS objects, and $0.06 < r < 10$ for PALFA objects. Our $(S/N_{\rm SP})/(S/N_{\rm FFT})$ values fall in the range quoted by \citet{Deneva:2009} for the PMPS, except for three instances where the values are slightly higher. The longer observation time of the PMPS (35 min) compared to PALFA (4.5 min) allows us to record a larger number of pulses. Based on fig. 12 from \citet{McLaughlin:2003}, larger values of $N_{\rm p}$ increase $(S/N_{\rm SP})/(S/N_{\rm FFT})$ if the pulse amplitudes follow a power-law with an index between $-1.5$ and $-3$, which one of these pulsars (J0828$-$3417) does (see Table \ref{table:energy}). The other two therefore are likely to show strong intrinsic modulation. Unlike \citet{McLaughlin:2003}, we find that $r$ increases for decreasing $N_{\rm p}$, for any range of power-law indices. \citet{Deneva:2009} saw this same behaviour, though they did not measure power-law indices.

\begin{figure*}
\centering
\includegraphics[height=\textwidth, angle=-90]{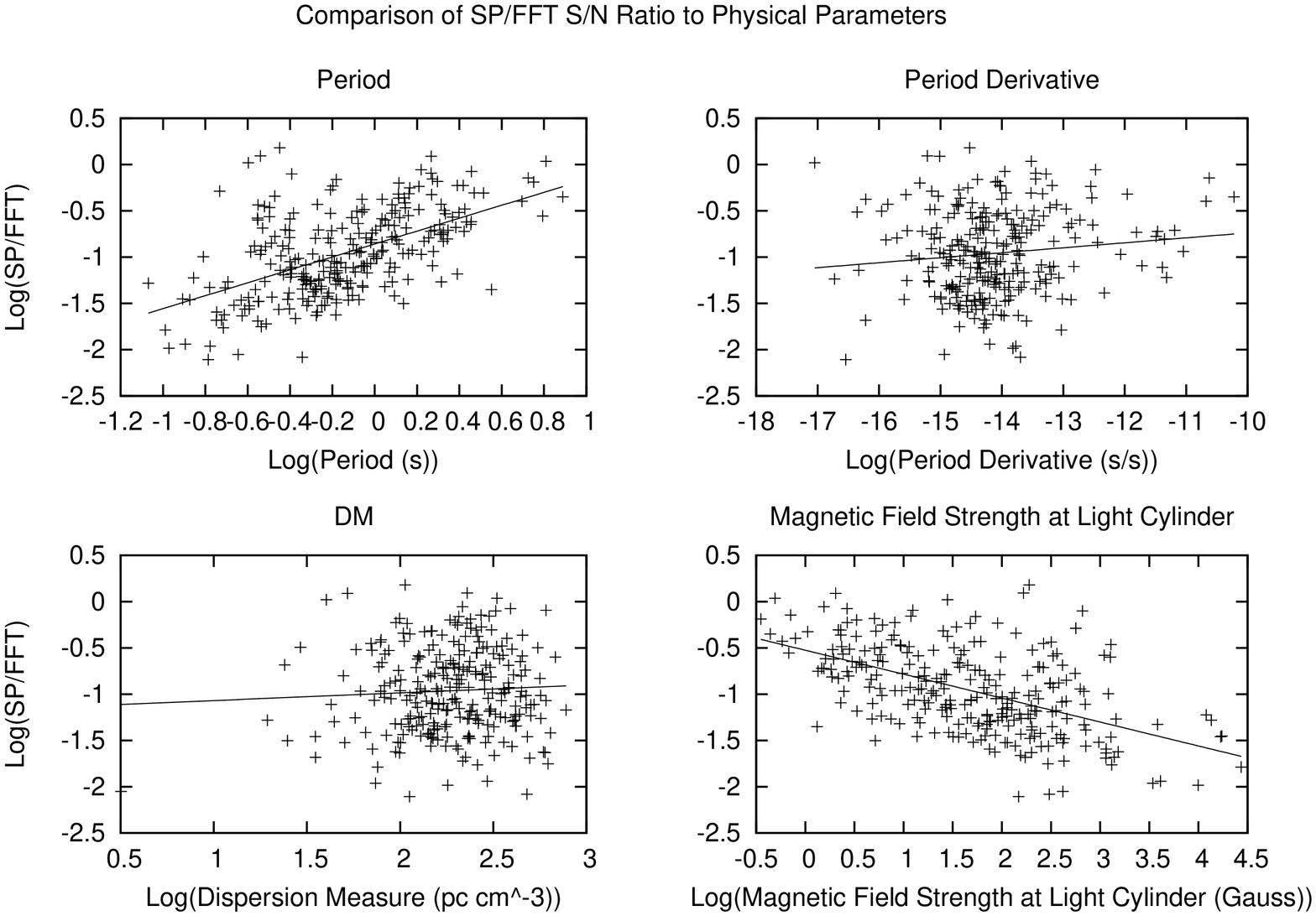}
\caption[SP/FFT S/N vs physical parameters]{Examples of SP/FFT S/N as a function of physical parameters. Clockwise from the top left are period, period derivative, magnetic field at the light cylinder, and DM.}
\label{fig:ratio_vs_params}
\end{figure*}

We compared this $(S/N_{\rm SP})/(S/N_{\rm FFT})$ ratio for each pulsar to that pulsar's physical properties (spin period, period derivative, DM, characteristic age, spin-down energy loss rate, surface magnetic field strength, and magnetic field strength at the light cylinder) by fitting a straight line to scatter plots of parameter values versus the S/N ratio. Examples of these fits can be seen in Fig. \ref{fig:ratio_vs_params}. The resulting correlations, including their $\chi_{\rm red}^2$ and significance, can be found in Table \ref{table:pmps_corrs}. We list a significance as the error on the slope of the fitted straight line. Also included in the Table are the Spearman $\rho$ rank correlation coefficients and p-values, where a p-value $> 0.1$ is consistent with zero correlation \citep{Press:1986}. As can be seen from Table \ref{table:pmps_corrs}, the spin period, characteristic age, spin-down energy loss rate, and the magnetic field strength at both the surface and the light cylinder have significant correlations, while the period derivative and DM show very weak correlations. We do not use the $\chi_{\rm red}^2$ as a determining factor here, since the large scatter in the correlations limits its usefulness. The Spearman correlation coefficient results reflect this.

\begin{table*}
\centering
\caption[Correlation Between $(S/N_{\rm SP})/(S/N_{\rm FFT})$ and Physical Characteristics]{Correlation between the ratio of maximum single-pulse S/N to FFT S/N and physical parameters of the pulsar, including the spin period, period derivative, DM, characteristic age, spin-down energy loss rate, magnetic field strength at the surface, and magnetic field strength at the light cylinder. From left to right we list the slope estimates, estimated standard deviations, and resulting $\chi_{\rm red}^2$ from linear fits to the data, as well as the significance of the slope. We also list the Spearman rank correlation coefficient and its p-value for a test of departure from zero.}
\begin{tabular}{cccccc}
\hline
\hline
 & \multicolumn{2}{c}{Fit Statistics} & Significance & Correlation & Correlation\\
\cline{2-3}
 & Slope & $\chi_{\rm red}^2$ & & Coefficient & p-value\\
\hline
$P$ & 0.70$\pm$0.07 & 0.15 & 10$\sigma$ & 0.54 & $7\times10^{-21}$\\
$\dot{P}$ & 0.05$\pm$0.03 & 0.22 & 1$\sigma$ & 0.10 & 0.11\\
DM & 0.08$\pm$0.09 & 0.22 & None & 0.03 & 0.67\\
$\tau_{\rm c}$ & 0.15$\pm$0.03 & 0.20 & 4$\sigma$ & 0.25 & $4\times10^{-5}$\\
$\dot{E}$ & $-$0.18$\pm$0.02 & 0.17 & 8$\sigma$ & $-$0.47 & $1\times10^{-15}$\\
$B_{\rm surf}$ & 0.23$\pm$0.06 & 0.21 & 3$\sigma$ & 0.21 & $6\times10^{-4}$\\
$B_{\rm LC}$ & $-$0.26$\pm$0.03 & 0.16 & 9$\sigma$ & $-$0.52 & $4\times10^{-19}$\\
\end{tabular}
\label{table:pmps_corrs}
\end{table*}

Period derivative and DM do not display significant correlations with 
$(S/N_{\rm SP})/(S/N_{\rm FFT})$. The spin period, however, is the most significant variable. We 
use this fact to explain the rest of the correlations. The characteristic age, spin-down energy 
loss rate, and the magnetic field strength at both the surface and the light cylinder are all 
strongly dependent on the spin period, as shown in \citet{Lorimer:2004} by the following scaling 
laws
\begin{equation}
\tau_{\rm c}\propto\frac{P}{\dot{P}},\;\;\dot{E}\propto\frac{\dot{P}}{P^3},\;\;B_{\rm surf}\propto\sqrt{P\dot{P}},\;\;B_{\rm LC}\propto\frac{\dot{P}^{1/2}}{P^{5/2}}.
\end{equation}
Hence, the strong correlations seen among these properties are dominated by the correlation with spin period. This correlation with spin period does not mean that longer period pulsars have brighter single pulses, however. Longer period pulsars have fewer pulses per observation than short period pulsars. According to fig. 12 of \citet{McLaughlin:2003}, for a sufficiently small number of pulses $N_{\rm p}$, you are more likely to detect a pulsar more strongly in single pulses than in a periodicity search for any value of the power-law index of its pulse amplitude distribution. A secondary factor contributing to decreased $S/N_{\rm FFT}$ in longer period pulsars could be the generally shorter duty cycle of long period pulsars, which may cause power to be lost if it spreads into higher harmonics.

As mentioned before, \citet{Burke-Spolaor:2012} studied similar correlations. They 
searched for correlations between physical properties and the R modulation statistic, defined by 
\citet{Johnston:2001}. Like us, they found several strong correlations, including those with 
characteristic age and spin-down energy loss rate, but these were all determined to be due 
to a strong anti-correlation between integrated S/N and the modulation index.

This work and \citet{Burke-Spolaor:2012} suggest that there is no physical meaning to the correlations between physical properties and $(S/N_{\rm SP})/(S/N_{\rm FFT})$; they are merely an observational effect. However, \citet{Cui:2017} compared cumulative probability distributions of RRATs to those of 
pulsars for distance, period, surface magnetic field strength, characteristic age, spin-down 
energy loss rate, and magnetic field at the light cylinder using a Kolmogorov-Smirnoff test 
\citep{Press:1986}. They found that the distributions of period derivatives for pulsars and 
RRATs are different, perhaps indicating a link between spin-down properties and emission properties.

\section{RRATs}
\label{rrats}

We based our analysis of single pulses from RRATs on the beam closest to their catalogued positions, except in the case of RRATs originally discovered by \citet{McLaughlin:2006}, where we instead base our analysis on the beam in which they were originally detected. We analysed those RRATS from which we detected single pulses when examining 
the single-pulse plot by eye. This led to a total of 14 RRATs being included in our analysis.
We created energy distributions in a similar manner to the way in which we created them for 
known pulsars\footnote{Plots of the fits to the RRAT energy distributions can be found at \url{http://astro.phys.wvu.edu/pmps/lognormal.html}}.

\begin{figure*}
\centering
\includegraphics[width=\textwidth]{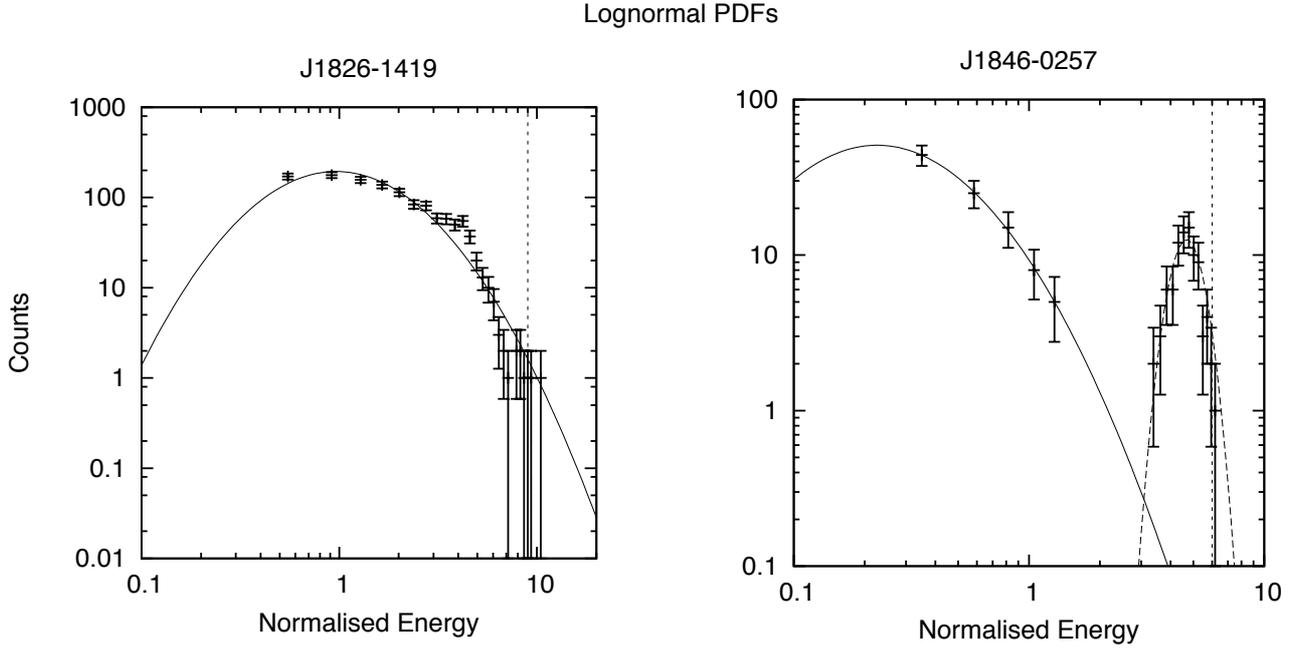}
\caption[Examples of Log-normal Fits to RRAT Amplitude Distributions]{Amplitude distributions for RRATs with log-normal fits. The plot on the left is one of the two examples of RRATs which do not exhibit `bumps' on their amplitude distributions, while the plot on the right shows this `bump', as do most of the RRATs we analysed. The RRATs are, from left to right, J1826$-$1419 and J1846$-$0257. The dashed vertical lines represent our detection thresholds.}
\label{fig:rrats}
\end{figure*}

The majority of the RRATs also show `bumps' in their distribution (see Fig. \ref{fig:rrats}), 
similar to the bursting behaviour seen in some of the known pulsars (see Fig. \ref{fig:burst}). However, most of the distributions drop nearly to zero between the peaks, resulting in two distinct distributions. For this reason, we attempted to fit two log-normal functions to each distribution. The results of these fits are listed in Table \ref{table:rrats}.

\begin{table*}
\centering
\small
\rotate
\caption[Single-pulse Detections of RRATs]{Single-pulse detections of RRATs from the reprocessing 
of the PMPS. From left to right we list the RRAT name, spin period, DM, number of single pulses 
detected above our threshold out of the total number of stellar rotations during the observation, best-fitting
parameters from log-normal fits and the $\chi_{\rm red}^2$ of the fit, and peak single-pulse S/N. The
fits were calculated through least-squares fitting of a scaled log-normal distribution, as above. 
The amplitude distributions for most RRATs are fit by two log-normal distributions, with the first 
distribution being consistent with noise, while the second distribution represents single pulses 
from the RRAT. Stars in the log-normal fit columns signify that we were unable to fit a log-normal 
function to that RRAT's amplitude distribution.}
\begin{tabular}{cccccccccccc}
\hline
\hline
RRAT & Period & DM & Number of Single & \multicolumn{3}{c}{First Log-normal} & & \multicolumn{3}{c}{Second Log-normal} & Max SP\\
\cline{5-7} \cline{9-11}
 & (s) & (pc cm$^{-3}$) & Pulses Detected & $\mu (E/\langle{E}\rangle)$ & $\sigma (E/\langle{E}\rangle)$ & $\chi_{\rm red}^2$ & & $\mu (E/\langle{E}\rangle)$ & $\sigma (E/\langle{E}\rangle)$ & $\chi_{\rm red}^2$ & S/N\\
\hline
J0847$-$4316 & 5.977 & 292.50 & 4/351 & $-$8.8(40.7) & 2.82(7.21) & 1.12 & & 1.53(0.01) & 0.11(0.01) & 0.52 & 10.77\\
J1047$-$58 & 1.231 & 69.30 & 17/1705 & * & * & * & & * & * & * & 23.65\\
J1317$-$5759 & 2.642 & 145.30 & 2/794 & $-$0.50(0.06) & 0.47(0.05) & 1.10 & & 1.18(0.02) & 0.21(0.02) & 1.01 & 24.79\\
J1444$-$6026 & 4.759 & 367.70 & 2/441 & $-$0.58(0.27) & 0.72(0.26) & 1.58 & & 1.60(0.02) & 0.20(0.02) & 0.38 & 42.58\\
J1724$-$35 & 1.422 & 554.90 & 3/1476 & 0.45(0.96) & 1.87(2.35) & 3.28 & & 1.47(0.03) & 0.27(0.03) & 1.08 & 12.69\\
J1754$-$3014 & 1.320 & 99.38 & 4/1590 & $-$0.10(0.11) & 0.69(0.16) & 2.54 & & 1.21(0.02) & 0.28(0.02) & 0.62 & 9.60\\
J1819$-$1458 & 4.263 & 196.00 & 11/492 & $-$1.67(0.99) & 1.58(0.57) & 0.08 & & 1.16(0.02) & 0.12(0.01) & 1.52 & 35.42\\
J1826$-$1419 & 0.771 & 160.00 & 2/2723 & 0.48(0.04) & 0.72(0.04) & 3.34 & & * & * & * & 26.77\\
J1839$-$0136 & 0.933 & 307.00 & 2/2250 & 0.50(0.05) & 0.82(0.04) & 2.18 & & * & * & * & 53.83\\
J1840$-$1419 & 6.598 & 19.40 & 69/318 & * & * & * & & 0.69(0.04) & 0.15(0.02) & 5.34 & 27.53\\
J1846$-$0257 & 4.477 & 237.00 & 2/469 & $-$0.84(0.05) & 0.81(0.06) & 0.03 & & 1.57(0.02) & 0.15(0.02) & 0.36 & 9.48\\
J1848$-$1247 & 0.414 & 88.00 & 3/5072 & * & * & * & & 1.299(0.006) & 0.135(0.008) & 0.05 & 10.91\\
J1911+0037 & 6.940 & 100.00 & 1/302 & * & * & * & & * & * & * & 9.23\\
J1913+1330 & 0.923 & 175.64 & 4/2275 & 0.22(0.02) & 0.72(0.03) & 0.59 & & 1.59(0.03) & 0.21(0.03) & 1.46 & 8.57\\
\end{tabular}
\label{table:rrats}
\end{table*}

In general, the log-normal fits for the first distribution are generally poor, with high errors. 
The mean values for many of these distributions are near zero, showing that they are consistent 
with noise. The fits for the second distributions, which we interpret to be actual RRAT 
pulses, have $\mu$ and $\sigma$ values that are consistent with those of the known 
pulsars we analysed, although the ranges of both $\mu$ and $\sigma$ are smaller.

Two of the RRATs analysed only have a first log-normal fit. One of these two is shown on the left of Fig. \ref{fig:rrats}. We interpret these two sources to be similar to normal pulsars, as energy from every rotation period shows underlying emission, as opposed to the example source on the right of the figure, where many rotation periods clearly show only noise, with no underlying emission. Folding the data from the beam used for the analysis of each RRAT and normal pulsar candidate showed no discernible pulse profiles.

\section{Conclusions}
\label{conc}

We performed single-pulse searches of the Parkes Multibeam Pulsar Survey (PMPS) out to a DM of 
5000~pc cm$^{-3}$ and with sensitivity to pulses with widths of up to one second. The 35-min pointings of the PMPS make it more sensitive to faint objects in periodicity searches than new surveys, e.g.\ the High Time Resolution Universe survey, and also allow us to detect more single pulses, putting better constraints on our amplitude distributions. We detected single pulses from 264 known pulsars, 
as well as 14 RRATs, in the survey. We created normalised energy distributions using the energy 
from each rotation of these objects and fit these with log-normal distributions and power-law 
tails. For known pulsars, there is a wide range in the $\mu$ parameter of the log-normal fits 
(0 $E/\langle{E}\rangle < \mu <$ 3 $E/\langle{E}\rangle$) and a smaller range of 
$\sigma$ values (0.2 $E/\langle{E}\rangle < \sigma <$ 1.5 $E/\langle{E}\rangle$). Both 
of these parameters have slightly wider ranges than those seen in a previous study by 
\citet{Burke-Spolaor:2012}. This could be due to the longer observation times of the PMPS, which 
allows us to collect more single pulses and thereby constrain the amplitude distributions better. 
The values of the slope of the power-law fits ranged from $-$13.7 to 2.2. We also found that the power-law/exponential function, developed upon inspection of the energy distributions, fit the data more closely, with 121 of the fits resulting in $\chi_{\rm red}^2$ values between 1 and 2. For these fits, the values of $b$ ranged from $-$1.4 to 2.83 and the values of $c$ ranged from $-$0.06 to 2.25. We also calculated the ratio of the maximum single-pulse S/N and the FFT S/N, with resulting values from 0.01 to 1.51. These values are consistent with estimates by \citet{Deneva:2009}, with a few outliers due to longer observation times. We compared these ratios to the spin period, period derivative, DM, characteristic age, spin-down energy loss rate, surface magnetic field strength, and magnetic field strength at the light cylinder of each pulsar. We found no significant correlation between the 
$(S/N_{\rm SP})/(S/N_{\rm FFT})$ ratio and the period derivative and DM, but found strong correlations 
with spin period, characteristic age, spin-down energy loss rate, and the magnetic field strength 
at the surface and light cylinder. The strongest correlation was between the $(S/N_{\rm SP})/(S/N_{\rm FFT})$ 
ratio and spin period. The other strong correlations were determined to be dominated by the spin 
period correlation. The strong correlation with spin period does not necessarily mean that longer period pulsars have brighter single pulses, but is an observational effect due to the smaller number of pulses recorded from longer period pulsars in a given observation time, as shown by \citet{McLaughlin:2003}. Previous studies found similar strong correlations between modulation and some physical parameters. Some were attributed to an anti-correlation between the modulation index and integrated S/N, but others (for RRATs, specifically) seem plausible.

For most of the RRATs, their energy distributions are similar to those for our bursting pulsars. 
However, the `bump' is much more pronounced, enough so to be a distinct distribution. Indeed, the 
values of $\mu$ for the lower-energy distributions are near zero, showing these distributions are consistent with noise. The values of $\mu$ and $\sigma$ for the higher-energy distributions are consistent with those from the known pulsars, but their range is smaller. Two RRATs also show underlying emission in every rotation, lacking the noise distribution seen in most RRATs analysed here, and thus resemble normal pulsars.

In conclusion, we have shown that pulsars show a wide range of amplitude and pulse-energy distributions and that nulling pulsars, giant-pulsing pulsars, bursting pulsars and RRATs are all part of the same pulsar intermittency spectrum. This indicates that the same physical mechanism is responsible for emission in all of these sources, but that there are other variables which determine the specific emission properties for a particular source. High signal-to-noise observations of single pulses from pulsars, accompanied by robust physical modelling, are necessary to continue to make further progress.

\section*{Acknowledgements}

This work was supported by NSF award number OIA-1458952. MAM and DRL are members of the NANOGrav Physics Frontiers Center which is supported by NSF award 1430284. The Parkes radio telescope is part of the Australia Telescope National Facility which is funded by the Australian Government for operation as a National Facility managed by CSIRO.


\begin{deluxetable}{ccccccccccccccccc}
\tabletypesize{\tiny}
\rotate
\tablecaption{Single-pulse detections of known pulsars from the reprocessing of the PMPS. From left to right we list the pulsar name, spin period, DM, number of single pulses recorded, best-fitting parameters from log-normal fits and the $\chi_{\rm red}^2$ of the fit, best-fitting parameters from power-law fits and the $\chi_{\rm red}^2$ of the fit, best-fitting parameters from the power-law/exponential fits and the $\chi_{\rm red}^2$ of the fit, peak single-pulse S/N, FFT S/N, and the ratio of single-pulse S/N to FFT S/N. The fits were calculated through least-squares fitting of a scaled log-normal distribution, given by
$\frac{C}{x\sigma\sqrt{2\pi}}{{\rm exp}^-\frac{(ln(x)-\mu)^2}{2\sigma^2}}$, a power-law distribution, given by $Cx^\alpha$, and a power-law/exponential distribution, given by $a\frac{x^b}{e^{cx}}$. A star in the final two columns means that the pulsar was not detected in a periodicity search. Stars in the fit data columns signify that we were not able to fit the respective function to that pulsar's amplitude distribution. Here we provide a sample of the data; the full Table is available online.}
\tablewidth{0pt}
\tablehead{\colhead{PSR} &
\colhead{Period} &
\colhead{DM} &
\colhead{Number of Single} & 
\multicolumn{3}{c}{Log-normal} & &
\multicolumn{2}{c}{Power-law} & &
\multicolumn{3}{c}{Power-law/Exponential} & 
\colhead{Max SP} &
\colhead{Max FFT} &
\colhead{SP/FFT} \\
\cline{5-7} \cline{9-10} \cline{12-14}
\colhead{} &
\colhead{(s)} &
\colhead{(pc cm$^{-3}$)} &
\colhead{Pulses Detected} &
\colhead{$\mu (E/\langle{E}\rangle)$} &
\colhead{$\sigma (E/\langle{E}\rangle)$} &
\colhead{$\chi_{\rm red}^2$} & &
\colhead{$\alpha$} &
\colhead{$\chi_{\rm red}^2$} & &
\colhead{b} &
\colhead{c} &
\colhead{$\chi_{\rm red}^2$} &
\colhead{S/N} &
\colhead{S/N} &
\colhead{} }
\startdata
J0729$-$1836 & 0.510 & 61.29 & 694/4117 & 0.86(0.03) & 0.59(0.02) & 1.55 & & $-$4.35(0.60) & 1.12  & & 1.10(0.23) & 0.84(0.07) & 1.36 & 19.7 & 182.0 & 0.11 \\
J0820$-$4114 & 0.545 & 113.40 & 6/3853 & * & * & * & & $-$7.72(0.70) & 1.72  & & * & * & * & 7.2 & 124.0 & 0.06 \\
J0828$-$3417 & 1.849 & 52.20 & 3/1135 & 0.29(0.05) & 0.79(0.05) & 1.76 & & $-$3.14(1.49) & 1.34  & & 0.76(0.29) & 1.14(0.19) & 2.30 & 10.3 & 8.4 & 1.23 \\
J0837$-$4135 & 0.752 & 147.29 & 1466/2792 & * & * & * & & $-$7.73(0.72) & 1.13  & & * & * & * & 29.8 & 868.7 & 0.03 \\
J0842$-$4851 & 0.644 & 196.85 & 200/3260 & 0.54(0.05) & 0.79(0.04) & 3.98 & & $-$6.04(1.60) & 1.14  & & 0.64(0.11) & 0.77(0.05) & 1.41 & 13.1 & 145.9 & 0.09 \\
J0846$-$3533 & 1.116 & 94.16 & 393/1881 & 0.60(0.06) & 0.82(0.05) & 2.39 & & $-$6.33(0.72) & 0.42  & & 0.48(0.16) & 0.66(0.06) & 1.18 & 12.5 & 260.7 & 0.05 \\
J0904$-$4246 & 0.965 & 145.80 & 133/2176 & 0.59(0.05) & 0.84(0.04) & 1.95 & & $-$5.43(0.89) & 0.57  & & 0.45(0.12) & 0.65(0.05) & 0.97 & 18.9 & 122.5 & 0.15 \\
J0907$-$5157 & 0.254 & 103.72 & 1566/8267 & * & * & * & & $-$10.64(1.13) & 2.13  & & * & * & * & 17.5 & 478.6 & 0.04 \\
J0908$-$4913 & 0.107 & 180.37 & 8120/19626 & 1.68(0.03) & 0.78(0.03) & 3.78 & & $-$4.62(1.85) & 1.02  & & 0.46(0.07) & 0.22(0.01) & 1.06 & 14.3 & 1376.1 & 0.01 \\
J0924$-$5302 & 0.746 & 152.90 & 63/2815 & 0.46(0.05) & 0.74(0.04) & 4.60 & & $-$5.96(1.45) & 0.63  & & 0.78(0.15) & 0.92(0.07) & 2.09 & 7.1 & 156.5 & 0.05 \\
\enddata
\label{table:energy}
\end{deluxetable}






\bsp    
\label{lastpage}
\end{document}